\documentclass[useAMS,usenatbib]{mn2e}
\usepackage{tikz}
\usepackage{amssymb}

\title[Giant planet formation in self-gravitating discs]{Exploring the conditions required to form giant planets via gravitational instability in massive protoplanetary discs}
\author[Farzana Meru and Matthew R. Bate]{Farzana Meru$^{1}$\thanks{farzana@astro.ex.ac.uk} and Matthew R. Bate$^{1}$\\
$^{1}$School of Physics, University of Exeter, Stocker Road, Exeter, EX4 4QL}
\begin{document}

\maketitle

\label{firstpage}

\begin{abstract}
We carry out global three-dimensional radiation hydrodynamical simulations of self-gravitating accretion discs to determine if, and under what conditions, a disc may fragment to form giant planets.  We explore the parameter space (in terms of the disc opacity, temperature and size) and include the effect of stellar irradiation.  We find that the disc opacity plays a vital role in determining whether a disc fragments.  Specifically, opacities that are smaller than interstellar Rosseland mean values promote fragmentation (even at small radii, $R < 25$AU) since low opacities allow a disc to cool quickly.  This may occur if a disc has a low metallicity or if grain growth has occurred.  With specific reference to the HR~8799 planetary system, given its star is metal-poor, our results suggest that the formation of its imaged planetary system could potentially have occurred by gravitational instability.  We also find that the presence of stellar irradiation generally acts to inhibit fragmentation (since the discs can only cool to the temperature defined by stellar irradiation).  However, fragmentation may occur if the irradiation is sufficiently weak that it allows the disc to attain a low Toomre stability parameter.
\end{abstract}


\section{Introduction}
\label{sec:intro}

There are two ways in which giant planets have been hypothesised to form: core accretion \citep{Safronov_CA,Goldreich_Ward_CA,Pollack_etal_CA} and gravitational instability (\citealt{GI_Cameron,Boss_GI}; review by \citealt{GI_review_Durisen_PPV}).  The former has been favoured but historically has had difficulties for two reasons: the first is a temporal issue where the timescales required to form planets may be too large such that the gas in the disc is depleted before the gas giant planet is fully formed.  Secondly, simulations of this method model the growth of planets typically starting with kilometre-sized planetesimals, but while the growth of particles from small grains to metre-sized objects appears to be straight-forward, how to get from metre-sized objects to kilometre-sized planetesimals is unknown.

Gravitational instability, on the other hand, eliminates the timescale problem, forming gas giant planets in $\lesssim$~$\rm O(10^4)$ years.  Such planets may not have solid cores, which may well be the case for Jupiter \citep{Saumon_Guillot2004}, though the capture of solid material to form a core after the fragmenting stage in a gravitationally unstable disc has also been proposed \citep*{Solid_capture_postGI}.  However, since gravitational instability is not thought to operate close to the central star, it has not been thought to be the dominant mechanism by which giant planets form as it was unable to describe the observations of close-in giant planets.  Recent observations of planets at large distances \citep{Fomalhaut,HR8799} encourage us to revisit this theory.  \cite{Nero_Bjorkman_GI_analysis} have argued analytically that Fomalhaut b and at least the outer planet of the HR~8799 system could have formed by gravitational instability as the cooling timescales are likely to be small enough such that fragmentation is possible.  Moreover, discs in their early stages are thought to be massive \citep{Eisner_Carpenter_massive_discs} suggesting that gravitational instability must play a role in the evolution of a disc in the late Class I and early Class II stages.  It has also been proposed that core accretion may be a method by which planets may form at small radii ($\sim \rm{O}(10) \rm{AU}$) whilst gravitational instability may be the mechanism by which planets may form at larger radii ($\gtrsim \rm{O}(100) \rm{AU}$) \citep[e.g.][]{Boley_CA_and_GI} though a hybrid scenario of forming gas giants in the same system by both core accretion and gravitational instability has yet to be modelled.

There are two quantities that have historically been used to determine whether a disc is likely to fragment.  The first is the stability parameter \citep{Toomre_stability1964},

\begin{equation}
  \label{eq:Toomre}
  Q=\frac{c_{\rm s}\kappa_{\rm ep}}{\pi\Sigma G},
\end{equation}
where $c_{\rm s}$ is the sound speed in the disc, $\kappa_{\rm ep}$ is the epicyclic frequency, which for Keplerian discs is approximately equal to the angular frequency, $\Omega$, $\Sigma$ is the surface mass density and $G$ is the gravitational constant.  Therefore, once the surface mass density and the rotation of the disc have been established, the stability is purely dependent on the disc temperature.  \cite{Toomre_stability1964} showed that for an infinitesimally thin disc to fragment, the stability parameter must be less than a critical value, $Q_{\rm crit} \approx 1$.

\cite{Gammie_betacool} showed that in addition to the stability criterion above, the disc must cool at a fast enough rate.  Using shearing sheet simulations, he showed that if the cooling timescale can be parametrized as

\begin{equation}
  \label{eq:beta}
  \beta = t_{\rm cool}\Omega,
\end{equation}
where

\begin{equation}
  \label{eq:tcool}
  t_{\rm cool} = u \big(\frac{du_{\rm cool}}{dt}\big)^{-1},
\end{equation}
$u$ is the internal energy and $du_{\rm cool}/dt$ is the total cooling rate, then for fragmentation we require $\beta \lesssim 3$, for a ratio of specific heats $\gamma = 2$ (in two dimensions).  \cite*{Rice_beta_condition} carried out three-dimensional simulations using a smoothed particle hydrodynamics (SPH) code and showed that this cooling parameter is dependent on the equation of state so that fragmentation can occur for discs with $\gamma = 5/3$ and $7/5$ if $\beta \lesssim 7$ and $13$, respectively.  \cite*{Libby_MSci} showed that the critical value of $\beta$ (below which fragmentation will occur if the stability criterion is met) may depend on the disc's thermal history: if the timescale on which the disc's cooling timescale is decreased is slower than the cooling timescale itself (i.e. a gradual decrease in $\beta$) then the critical value may decrease by up to a factor of 2.  More recently, \cite*{Cossins_opacity_beta} showed that the critical value varies with the temperature dependence of the cooling law.  In general though, the critical value is thought to be of the order of the dynamical timescale.

The above fragmentation criteria are based on the assumption that the dominant form of dissipation in the disc is due to internal heating processes.  Previous simulations without external irradiation have considered isolated discs with simple cooling prescriptions \citep[e.g.][]{Lodato_Rice_original} and with radiative transfer \citep[e.g.][]{Boss_RT,Cai_etal_RT,Mayer_etal_RT}.  \cite{Johnson_Gammie} suggested that discs with external irradiation are likely to be effectively isothermal and can therefore be treated as such.  \cite{Matzner_Levin2005} analytically considered externally irradiated discs and concluded that stellar irradiation quenches fragmentation.  \cite{Cai_envelope_irradiation} carried out simulations with external irradiation and found that their discs are more resistant to fragmentation and proposed that these results may be extended to discs with stellar irradiation.   \cite{Stamatellos_no_frag_inside_40AU} also carried out simulations taking into account the effects of stellar irradiation and found this to be a stabilising factor.  \cite{Rafikov_SI} analytically explored fragmentation in gravitoturbulent discs including the effects of stellar irradiation and suggested that fragmentation can only occur beyond $\approx 120\rm{AU}$.  \cite{Dodson-Robinson_HR8799} carried out a linear stability analysis on irradiated discs to show that gravitational instability takes place for systems with a large disc to star mass ratio.  However, whilst fragmentation in gravitationally unstable discs may be less likely than previously thought, we still do not know in \emph{what situations} discs may fragment when modelling them realistically with radiative transfer and by considering the effects of stellar irradiation.  It is therefore important to deduce when fragmentation may occur when simulating discs with more detailed energetic conditions, and just how realistic or unrealistic fragmentation is in real discs.

\cite{Boss_metallicity} carried out simulations of gravitationally unstable discs and varied the opacity from $0.1\times$ to $10\times$ the Rosseland mean opacities and found that the fragmentation results were insensitive to the dust grain opacity.  However, given that a reduced opacity is more likely to allow energy to stream out of a disc more easily causing it to cool and promote fragmentation, whilst in a high opacity disc the converse is true, it is interesting to consider what opacity values allow and do not allow fragmentation.  Given that a disc's opacity gives somewhat an indication of how metal-rich it is or how large or small the grain sizes are, we may then make preliminary conclusions on the disc conditions that are likely to promote fragmentation, which is a key focus of this paper.

In this paper, we model the evolution of massive self-gravitating discs using a global three-dimensional SPH code including radiative transfer and the effects of stellar irradiation.  In particular, we explore the parameter space in terms of dust opacity, disc temperature and size in order to scope out if, and under what conditions, a self-gravitating disc may fragment.  In Section~\ref{sec:numerical_setup}, we describe the code used to carry out our simulations.  In Section~\ref{sec:sim}, we outline our simulations, including the disc setup and discuss the parameter space.  In Section~\ref{sec:results}, we present our results, while we compare with previous studies and make conclusions in Sections~\ref{sec:disc} and~\ref{sec:conc}, respectively.

\section{Numerical setup}
\label{sec:numerical_setup}

Our simulations are carried out using an SPH code originally developed by \cite{Benz1990} and further developed by \cite*{Bate_Bonnell_Price_sink_ptcls}, \cite*{WH_Bate_Monaghan2005}, \cite{WH_Bate_science} and \cite{Price_Bate_MHD_h}.  It is a Lagrangian hydrodynamics code, ideal for simulations that require a large range of densities to be followed, such as fragmentation scenarios.  The version used to carry out the simulations presented here includes radiative transfer using the flux-limited diffusion approximation \citep{WH_Bate_Monaghan2005,WH_Bate_science} with two temperatures: that of the gas and that of the radiation field.

In order to model shocks, SPH requires artificial viscosity.  We use a common form of artificial viscosity by \cite{Monaghan_Gingold_art_vis}, which uses the parameters $\alpha_{\rm SPH}$ and $\beta_{\rm SPH}$.  A corollary of including artificial viscosity is that it adds shear viscosity and causes dissipation.  If this viscosity is too large, the evolution of the disc may be driven artificially, while if it is too small, it will lead to inaccurate modelling of shocks \citep{Matthews_thesis}.  We have tested various values of the SPH parameters and find that a value of $\alpha_{\rm SPH} \sim 0.1$ provides a good compromise between these factors.  Since typically, $\beta_{\rm SPH} \sim 2\alpha_{\rm SPH}$, we choose $\alpha_{\rm SPH}$ and $\beta_{\rm SPH}$ to be 0.1 and 0.2, respectively.  These are the same values as those implemented by \cite{Lodato_Rice_original}.  However, we have chosen to fix our artificial viscosity parameters whereas \cite{Lodato_Rice_original} used the \cite{Balsara_switch} switch which reduces the artificial viscosity in places where the ratio of the divergence to the curl of the velocity field is small i.e. in regions of strong vorticity.  The reasoning behind our choice is two-fold: (i) the Balsara switch reduces the viscosity in shearing flows but given that we have both shocks and shearing flows, the reduction does not handle the shocks well; (ii) we want to ensure a controlled test which is not possible if the numerical viscosity is constantly changing.

\subsection{Opacity \& equation of state}
\label{sec:eos}

The discs simulated are assumed to be in local thermal equilibrium and the opacities are assumed to be grey Rosseland mean values.  The opacities are based on the opacity tables of interstellar molecular dust grains produced by \cite*{Pollack_etal_dust_opacitites}, and on \cite{Alexander_gas_opacities} for the higher temperature gaseous contributions.  A summary of how the opacity tables were used is available in \cite{WH_Bate_science}.  For non-interstellar opacity simulations, we scale these values by the required factor (see Table~\ref{tab:sim} for simulation details).  The details of how this was done can be found in Section 2.4 of \cite{Ayliffe_Bate1}.  

The equation of state used in these simulations assumes that the gas is composed of hydrogen (70\%) and helium (28\%) and includes dissociation of molecular hydrogen, ionisation of both hydrogen and helium, and the rotational and vibrational modes of molecular hydrogen, as done so by \cite{WH_Bate_science}, though the equation of state has been corrected following \cite{ortho_para_Boley}.  It omits the contribution due to metals.  Our discs are cold enough such that the rotational and vibrational degrees of freedom are not excited so that effectively, $\gamma = 5/3$.  Our models  assume that the presence of dust only affects the value of the opacity in the disc.  We do not include the effects of dust on the disc dynamics, nor do we consider its effects on the equation of state.

\subsection{Stellar irradiation}
\label{sec:SI}

As mentioned earlier, radiative transfer is simulated using the flux-limited diffusion approximation.  We use a two-layer approach to simulate the midplane and surface regions.  In order to model the energy loss from the disc surface, a boundary layer of particles maintains a fixed temperature profile such that any energy that is passed to these boundary particles is effectively radiated away.  The vertical location of the boundary between the optically thick part of the disc and the optically thin atmosphere is located at the maximum of: the height above the midplane where the optical depth, $\tau = 1$, or the height above the midplane where $z_{\rm b} = 1.75 H$, where $H$ is the isothermal scale height given by $H = c_{\rm s}/\Omega$.  Therefore, the boundary is at the vertical position where the optical depth, $\tau \lesssim 1$, and our choice ensures that the ``bulk'' of the disc (i.e. consisting of particles that lie closer to the disc midplane) is simulated using radiative transfer, rather than simplified energetic calculations.  The remaining ``boundary'' particles have their temperature held at a constant temperature profile.  Physically, we assume that the atmosphere temperature is not set by the disc itself but by some other process such as stellar or external irradiation.  The particles forming the disc boundary are variable since some particles may move across the boundary from the optically thin to the optically thick region and vice versa.  We have carried out tests to see what the effects of the exact location of the vertical boundary has on the disc energetics and have found that provided the boundary is located in the optically thin surface region of the disc and provided that a layer of boundary particles exists over the entire disc surface, the disc energetics remain unchanged.

\section{Simulations}
\label{sec:sim}

\begin{table*}
  {\footnotesize
  \centering
  \begin{tabular}{llllllll}
    \hline
    Simulation & Disc radius & Opacity & $Q_{\rm min}$ & Fragment? & Fragmentation\\
    name & [AU] & scaling factor & & & time\\
    \hline
    \hline
    Reference & 25 & 1 & 2 & n & - \\
    Kappa10 & 25 & 10 & 2 & n & - \\
    Kappa0.1 & 25 & 0.1 & 2 & n & - \\
    Kappa0.01 & 25 & 0.01 & 2 & n & - \\
    Qmin1 & 25 & 1 & 1 & n & - \\
    Qmin0.75 & 25 & 1 & 0.75 & n & - \\
    Qmin0.75-Kappa0.1 & 25 & 0.1 & 0.75 & n & - \\
    Qmin0.75-Kappa0.01 & 25 & 0.01 & 0.75 & y & 9.7 ORPs \\
    Qmin0.5 & 25 & 1 & 0.5 & y & 1.8 ORPs\\
    L-Qmin1 & 300 & 1 & 1 & n & - \\
    L-Qmin1-Kappa10 & 300 & 10 & 1 & n & - \\ 
    L-Qmin1-Kappa0.1 & 300 & 0.1 & 1 & y & 1.5 ORPs \\
    L-Qmin0.75 & 300 & 1 & 0.75 & y & $< 1$ ORPs \\
    \hline
  \end{tabular}
  }
  \caption{Summary of the simulations carried out.  The opacity scalings refer to multiples of interstellar Rosseland mean opacity values as described in Section~\ref{sec:numerical_setup}.  $Q_{\rm min}$ refers to the minimum value of the Toomre parameter (at the outer edge of the disc) at the start of the simulation.}
  \label{tab:sim}
\end{table*}

Table~\ref{tab:sim} summarises the parameters and fragmentation results of the simulations presented here.  Each simulation was run either beyond the point at which the disc attained a steady state (for $> 6$ outer rotation periods, ORPs), or until it fragmented (defined as regions which are at least three orders of magnitude denser than their surroundings).  The magnitude and profile of the boundary temperatures are the same as those of the initial discs so that the discs start in thermal equilibrium with their vertical boundaries (atmospheres).  We expect the discs to heat up initially due to work done and viscous heating.  Following an initial transient phase, the bulk of the disc may or may not cool to re-establish thermal equilibrium with the boundary.  We discuss a Reference case first before turning our attention to exploring the parameter space.

\subsection{Reference case}
\label{sec:ref}

Our reference disc is set up in exactly the same way as a disc simulated by \cite{Lodato_Rice_original}: a 1~$\rm{M_\odot}$ star with a 0.1~$\rm{M_\odot}$ disc made of 250,000 SPH particles, spanning $0.25 \le R \le 25$AU.  The initial surface mass density and temperature profiles of the disc are $\Sigma \propto R^{-1}$ and $T \propto R^{-\frac{1}{2}}$, respectively.  The magnitudes of these are set such that the Toomre stability parameter (equation~\ref{eq:Toomre}) at the outer edge of the disc, $Q_{\rm min} = 2$.  This gives an aspect ratio, $H/R \sim 0.05$.  We model the $1~\rm{M_\odot}$ star in the centre of the disc using a sink particle \citep{Bate_Bonnell_Price_sink_ptcls}.  At the inner disc boundary, particles are accreted onto the star if they move within a radius of 0.025~AU of the star or if they move into $0.025 \le R < 0.25 \rm{AU}$ and are gravitationally bound to the star.  At the outer edge, the disc is free to expand.

The initial Reference disc is a Toomre stable disc.  Given that the boundary temperature profile is the same as that of the initial disc and that $Q_{\rm min}=~2$, we do not expect this disc to fragment.  However, we use the Reference disc as a fiducial case.  In particular, we are concerned with the cooling rates that are present in the disc once it is in an equilibrium state.  We emphasize our use of terminology here: when referring to the disc being in \emph{thermal equilibrium with the boundary}, we are describing the bulk of the disc being a similar temperature to the disc boundary (which is assumed to be determined by stellar irradiation), whereas an \emph{equilibrium state} disc refers to the dissipative and cooling rates being balanced such that the Toomre stability profiles do not change with time.

\subsection{Exploring the parameter space}
\label{sec:parameter}
Given that a motivation of this work is to determine if, and under what circumstances, fragmentation in realistically modelled self-gravitating discs may occur, we explore the parameter space in a number of ways.  One parameter is the opacity: we re-run the Reference simulation with opacity values scaled to $10\times$, $0.1\times$ and $0.01\times$ the interstellar opacity values (simulations Kappa10, Kappa0.1 and Kappa0.01, respectively).  This may be equivalent to a disc with differing metallicities or grain sizes.  We make two assumptions here: i) the change in metallicity does not affect the equation of state (as described in Section~\ref{sec:eos}) and ii) there are no spatial or temporal variations in the grain size distributions.  As with the Reference case, these discs are simulated purely to analyse the energetics since we do not expect these discs to fragment.

We then choose to explore the initial and boundary temperature conditions by decreasing the magnitude of the disc temperature whilst maintaining the same surface mass density as the Reference case.  We do this by changing the initial Toomre stability parameter profiles such that $Q_{\rm min} = 1, 0.75$ and $0.5$ (simulations Qmin1, Qmin0.75 and Qmin0.5, respectively).  This is equivalent to reducing the disc aspect ratios to $H/R \sim 2.2 \times 10^{-2}$, $1.7 \times 10^{-2}$ and $1.1 \times 10^{-2}$, respectively.  We reiterate that the boundary temperature is the same as the temperature of the initial disc and hence this setup not only changes the disc temperature profile, but it also changes the boundary temperature profile.

Furthermore, we consider a combination of the above factors by simulating discs with $Q_{\rm min} = 0.75$ and opacities that are $0.1\times$ and $0.01\times$ the interstellar opacity values.
 
The unfavourable conditions for fragmentation at small radii have been discussed at great length in the past \citep[e.g.][]{Rafikov_unrealistic_conditions, Stamatellos_no_frag_inside_40AU, Boley_CA_and_GI,Rafikov_SI,Clarke2009_analytical}.  We therefore expand our parameter space to include discs that are a factor of 12 larger with a radii range of $3 \le R \le 300 \rm AU$.  These discs have the same mass as the 25AU discs and are set up so that $Q_{\rm min}=1$.  We simulate three different opacity values ($1\times, 10\times$ and $0.1\times$ the interstellar Rosseland mean opacities).  In addition, we also simulate a large disc with $Q_{\rm min} = 0.75$ with interstellar opacity values.

In order to keep these disc masses and initial Toomre stability profiles the same as the smaller 25AU discs, we require both the surface mass density and absolute temperature to be reduced.  These discs are therefore not only larger, but also colder than their equivalent (in terms of initial Toomre stability profiles) small discs.

\section{Results}
\label{sec:results}

The simulations have been analysed in three main ways:

(i) we compare the azimuthally averaged Toomre stability profiles of the initial and final (or in the case of fragmenting discs, shortly before fragmentation) discs which indicates whether the bulk of the discs were able to reach a state of thermal equilibrium with their boundary.  The surface mass density does not change significantly throughout the simulations and hence changes in the Toomre stability parameter are due to changes in the disc temperature.  This enables us to determine which discs are more likely to fragment.  Note that we assume $\kappa_{\rm ep} = \Omega$ in equation~\ref{eq:Toomre}.

(ii) we examine the timescale on which the discs cool (by considering the energy passed from the gas to the radiation within the disc as well as that which is assumed to be instantly radiated away from the disc surface by the boundary particles).  In past simulations that have neglected the heating effects of stellar irradiation \citep[e.g.][]{Gammie_betacool,Rice_beta_condition}, the cooling, $C$, in a steady-state disc balances the heating due to gravitational stresses, $H_{\rm GI}$, and the heating due to artificial viscosity, $H_{\rm \nu}$, such that

\begin{equation}
  \label{eq:cooling}
  C = H_{\rm GI} + H_{\rm \nu}.
\end{equation}
If the artificial viscosity is low, $C \approx H_{\rm GI}$.  In this case, the cooling timescale in units of the orbital timescale, $\beta$ (section~\ref{sec:intro}), can be related to the gravitational stress in the disc, $\alpha_{GI}$ \citep{Gammie_betacool}:

\begin{equation}
  \label{eq:beta_alpha}
  \alpha_{GI} = \frac{4}{9} \frac{1}{\gamma (\gamma - 1)} \frac{1}{\beta}.
\end{equation}  

\cite{Gammie_betacool} and \cite{Rice_beta_condition} have shown that the maximum gravitational stress that a disc can support is $\alpha_{GI} = 0.06$, beyond which fragmentation will occur.  In discs that do not take into account heating due to external irradiation, this condition is equivalent to requiring the cooling timescale in terms of the orbital timescale, $\beta$, to be smaller than the critical values, described in section~\ref{sec:intro}, for fragmentation.

In our steady-state discs, not only does the cooling have to balance the heating due to the gravitational instabilities and the numerical viscosity, but it also has to balance the heating due to stellar irradiation, $H_{SI}$, such that:

\begin{equation}
  \label{eq:cooling_SI}
  C = H_{\rm GI} + H_{\rm \nu} + H_{\rm SI}.
\end{equation}
In what follows, we calculate the parameter, $\psi$, which we define to be the timescale on which the disc cools in units of the orbital timescale.  Without irradiation, $\psi = \beta$.  When including heating due to stellar irradiation, the $\psi$~parameter does not specifically tell us about the fraction of the cooling that balances the gravitational instabilities and therefore cannot be used directly to decide whether a disc should fragment or not.  However, it \emph{does} still give an indication as to what the cooling rate is in the discs which has been shown to be important when determining whether a disc is likely to fragment or not \citep{Gammie_betacool,Rice_beta_condition}.

(iii) we examine images of the discs for signs of fragmentation which we define as clumps in the disc which are at least three orders of magnitude denser than their surroundings.

\subsection{Reference case}
\label{subsec:Ref}

\begin{figure}
  \includegraphics[width=1.0\columnwidth]{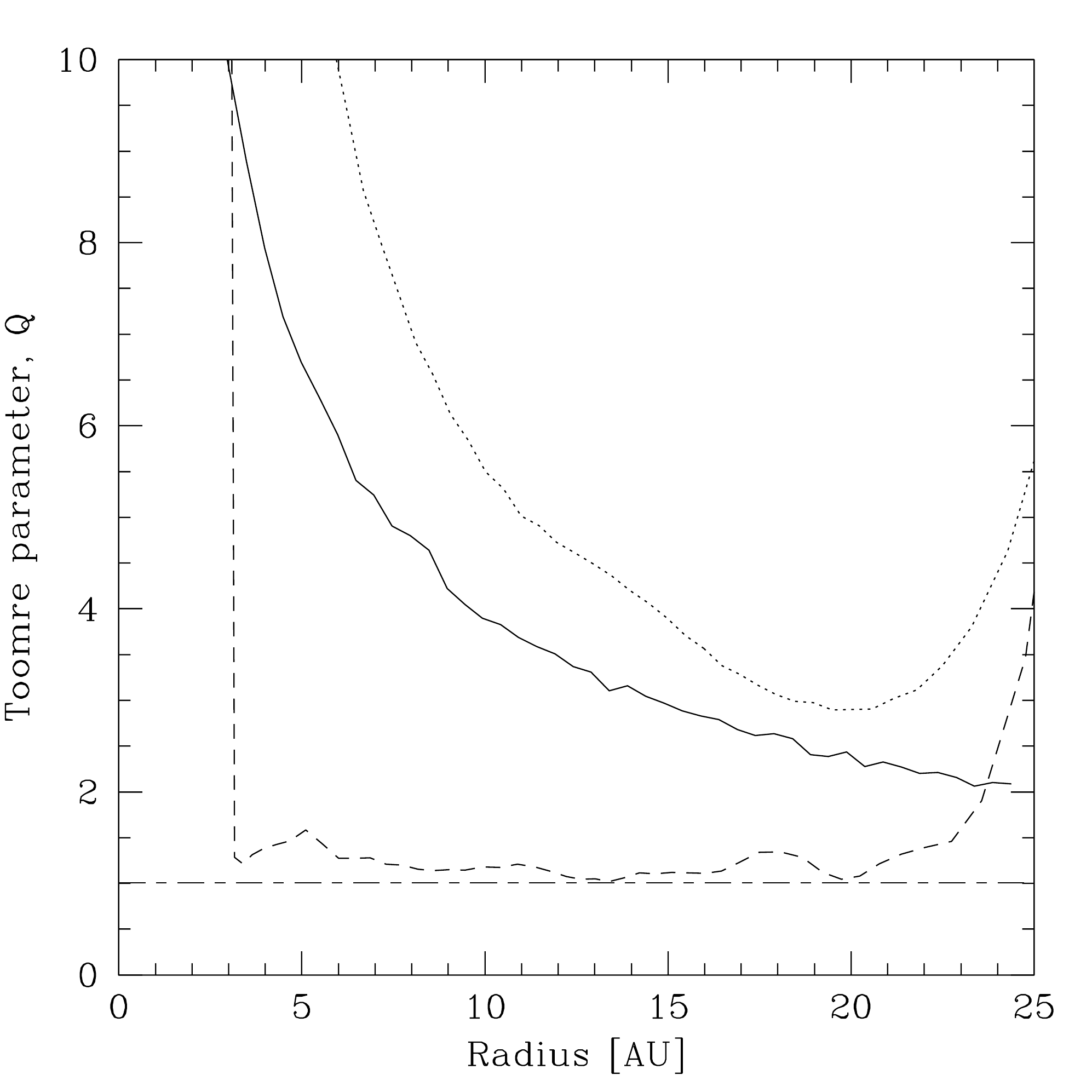}
  \caption{Azimuthally averaged values of the Toomre parameter at the start (solid line) and at a time $t = 6.4$ ORPs (dotted line) for the Reference simulation.  The disc is unable to cool rapidly due to the internal heating and hence its end state is more stable than the initial disc.  Also shown is the equivalent disc simulated by \citet[][short dashed line]{Lodato_Rice_original} which cools using simplified cooling (with $\beta=7.5$) rather than by radiative cooling, and also does not consider the effects of stellar irradiation.  The critical value of $Q_{\rm crit} = 1$ is also marked.}
  \label{fig:Q_Ref_start_end}
\end{figure}

\begin{figure}
  \includegraphics[width=0.55\columnwidth,angle=270]{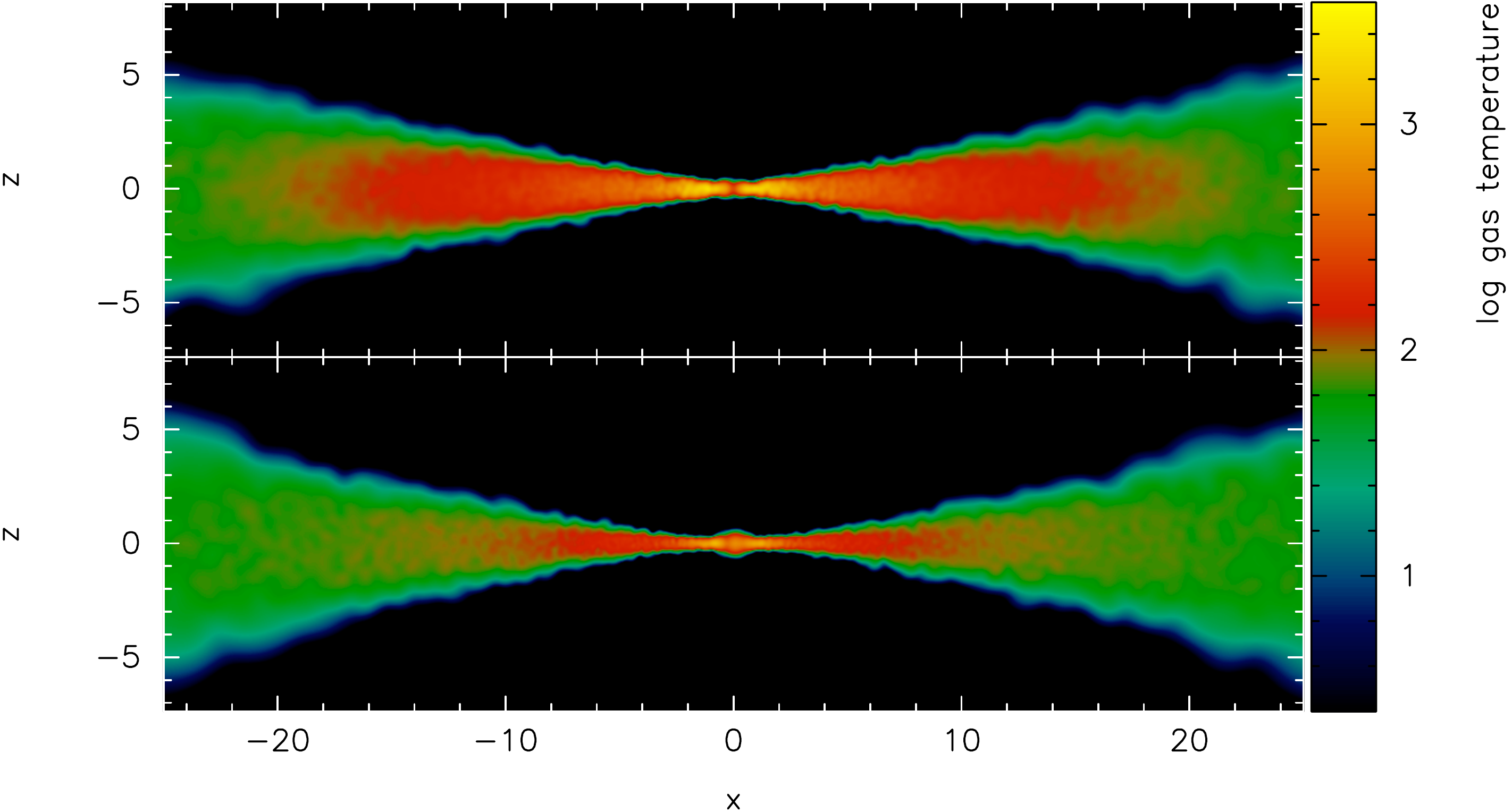}
    \caption{Logarithm of the gas temperature (in K) rendered in cross-sectional views of the Reference (top panel) and Kappa0.01 (bottom panel) discs at a time $t = 6.4$ ORPs.  The surface of the Reference disc is clearly colder than the midplane whilst the midplane of the Kappa0.01 disc is closer to a state of thermal equilibrium with the boundary.  Axis units are in AU.}
    \label{fig:Temp_rend}
\end{figure}

Figure~\ref{fig:Q_Ref_start_end} shows the Toomre stability profile of the initial and final Reference disc.  The disc is not able to cool rapidly enough in response to the internal heating and consequently the disc midplane becomes hotter than the boundary (top panel of Figure~\ref{fig:Temp_rend}).  Figure~\ref{fig:Q_Ref_start_end} also shows the final Toomre stability profile for the equivalent disc simulated by \cite{Lodato_Rice_original} which used simplified cooling rather than radiative cooling.  We have included this comparison to emphasize how significant the differences can be between discs simulated using simplified cooling parameters and discs modelled not only with more detailed radiative cooling but also incorporating the effects of stellar irradiation.

\subsection{Opacity effects}

\begin{figure}
  \includegraphics[width=1.0\columnwidth]{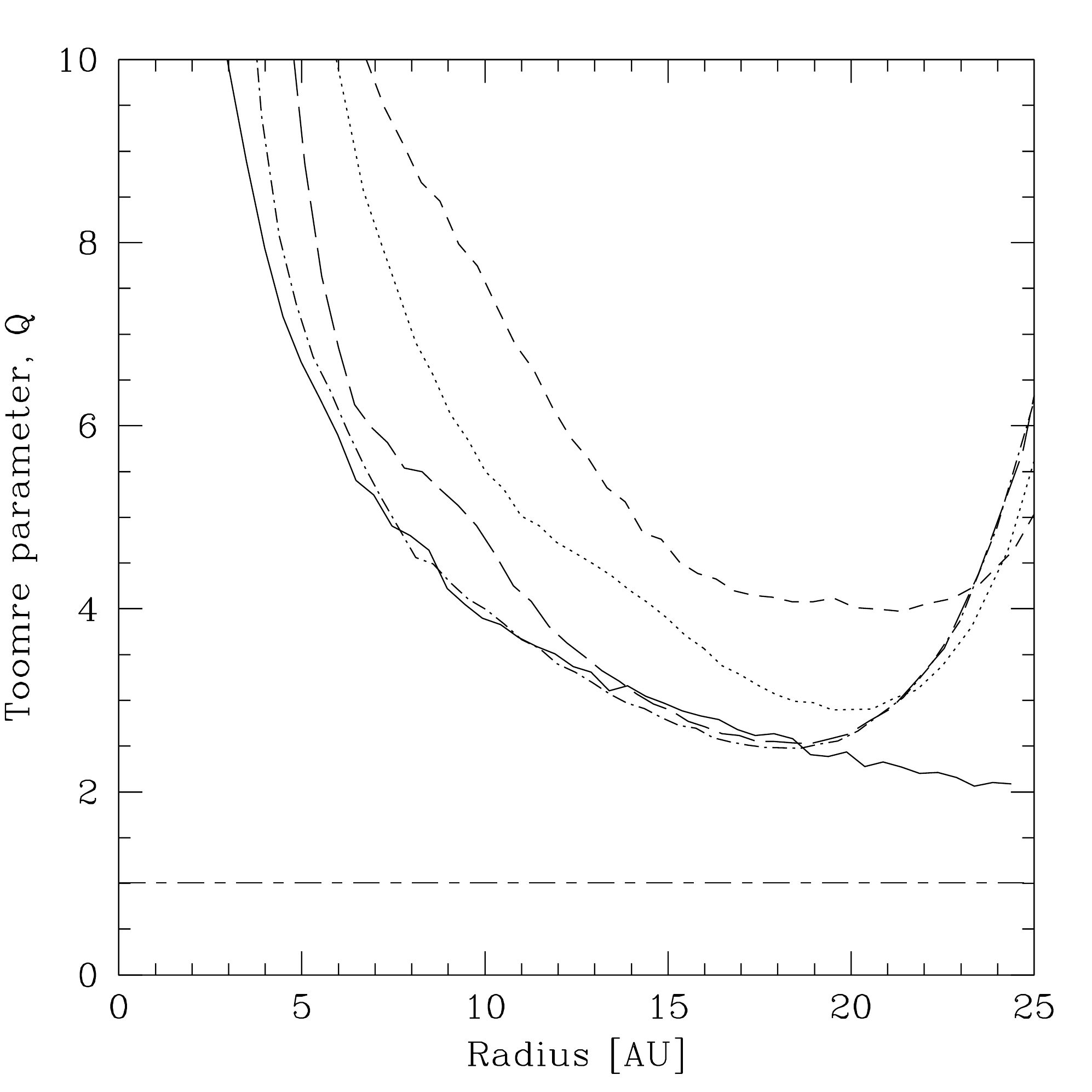}
  \caption{Azimuthally averaged values of the Toomre stability parameter at the start (solid line) and at a time $t = 6.4$~ORPs, for the Reference (dotted line), Kappa10 (short dashed line), Kappa0.1 (long dashed line) and Kappa0.01 (dot-dashed line) simulations.  The critical value of $Q_{\rm crit} = 1$ is also marked.  Decreasing the opacity causes the disc to cool more efficiently.}
  \label{fig:Q_kappa_end}
\end{figure}

\begin{figure}
  \includegraphics[width=1.0\columnwidth]{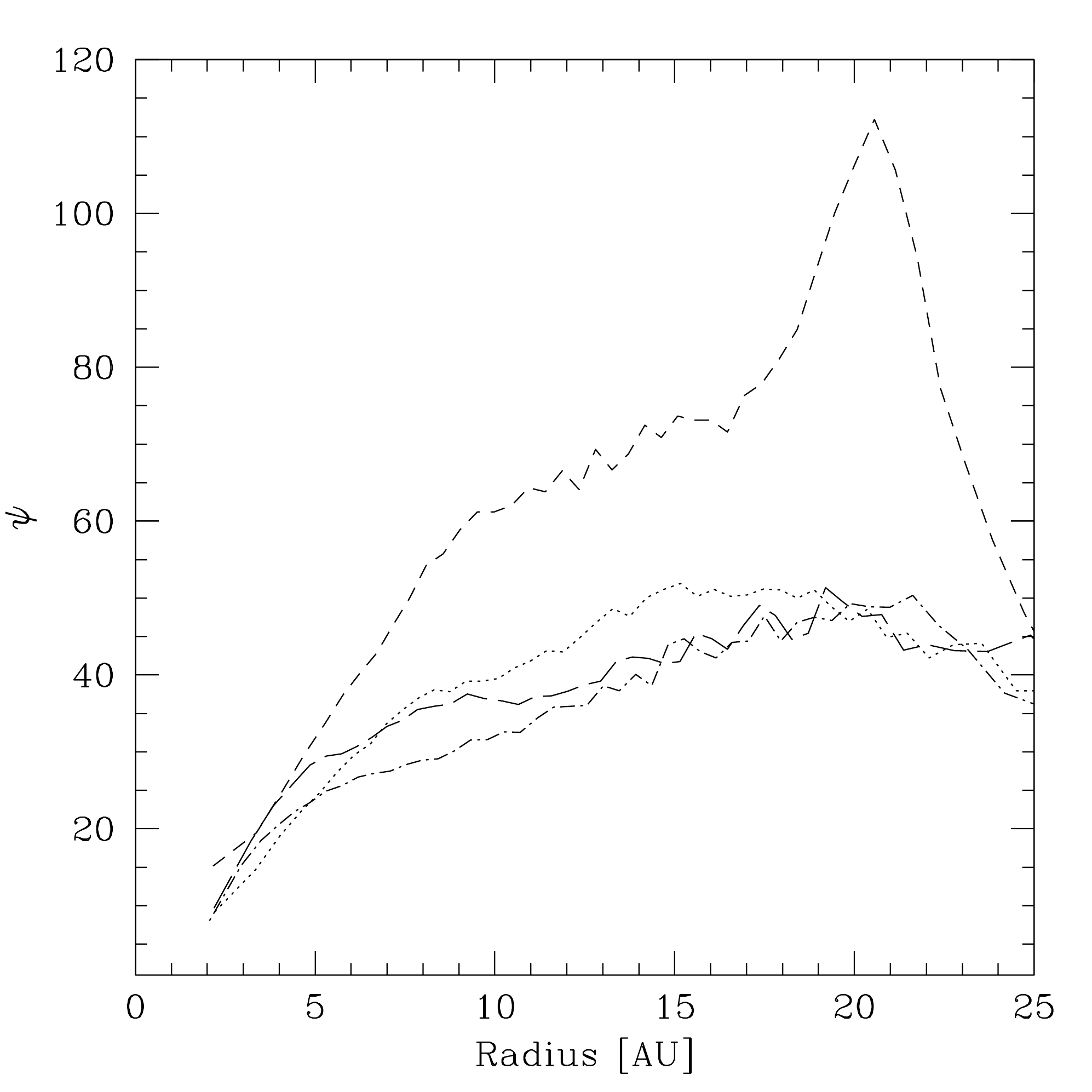}
  \caption{Cooling timescale, $\psi$, profiles for the Reference (dotted line), Kappa10 (short dashed line), Kappa0.1 (long dashed line) and Kappa0.01 (dot-dashed line) simulations at a time $t = 6.4$ ORPs.  Decreasing the opacity causes the cooling rate in the disc to increase (and $\psi$ to decrease), but only until the disc can reach thermal equilibrium with its boundary.}
  \label{fig:Beta_kappa}
\end{figure}

Figure~\ref{fig:Q_kappa_end} shows the effect on the Toomre stability parameter of changing the opacity in the discs to $10\times$, $0.1\times$ and $0.01\times$ the interstellar Rosseland mean values (simulations Kappa10, Kappa0.1 and Kappa0.01, respectively).  We see that decreasing the disc's opacity enhances its ability to reach a state of thermal equilibrium with the boundary since the radiation leaves the disc far more effectively and hence the disc cools faster.  This is particularly evident in the cross-sectional plots showing the temperature structure in the vertical direction (Figure~\ref{fig:Temp_rend}).  Figure~\ref{fig:Beta_kappa} shows the $\psi$ profile of these simulations.  This figure shows that a low opacity disc has a greater ability to radiate away the disc's energy.  The decrease in the cooling timescale, $\psi$, does not continue at very low opacities because the stellar irradiation sets the boundary temperature and therefore, the minimum disc temperature.

Figures~\ref{fig:Q_kappa_end} and \ref{fig:Beta_kappa} particularly show that the low opacity discs are able to cool fast enough to reach a state of thermal equilibrium with their boundaries.  Thus, if the conditions were right (i.e. the boundary temperature was lower so that the Toomre stability parameter was able to reach $Q \lesssim 1$), the low opacity disc may fragment.  We therefore turn our attention to the disc absolute temperature.

\subsection{Colder discs}

\begin{figure}
  \includegraphics[width=1.0\columnwidth]{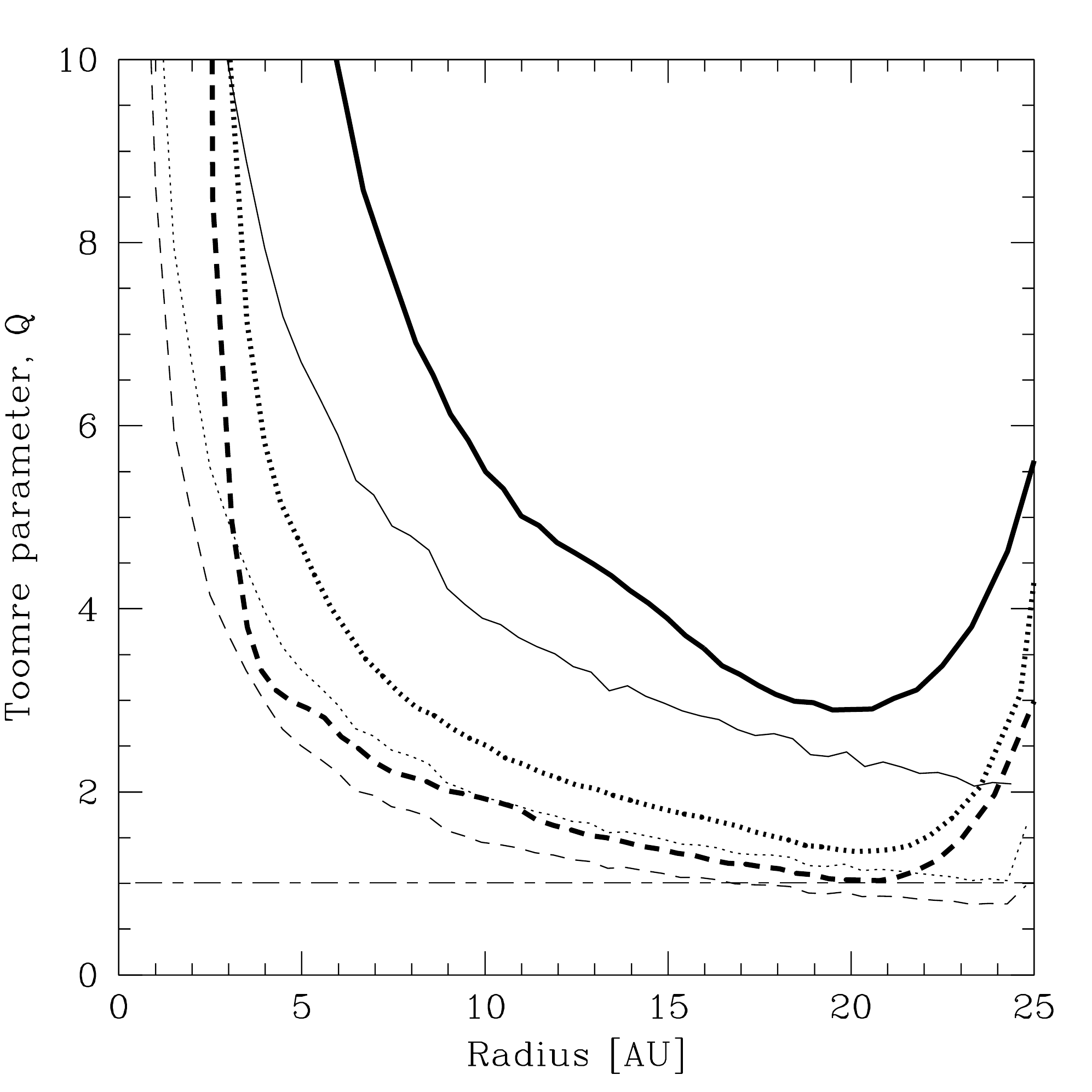}
  \caption{Initial (thin lines) and final (at $t = 6.4$ ORPs; heavy lines) Toomre stability profiles for the Reference (solid line), Qmin1 (dotted line) and Qmin0.75 (dashed line) simulations.  The critical value of $Q_{\rm crit} = 1$ is also marked.  None of the three discs which have interstellar opacity values can cool rapidly enough to maintain thermal equilibrium with their boundaries.}
  \label{fig:Q_Qmin_start_end}
\end{figure}

\begin{figure}
  \includegraphics[width=1.0\columnwidth]{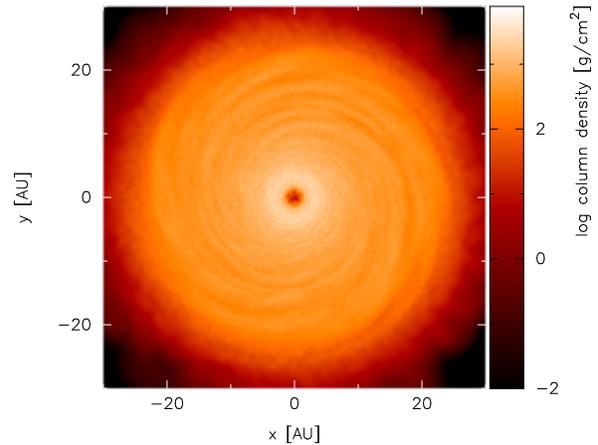}
  \caption{Surface density rendered image of the Qmin0.75 disc at a time $t = 6.4$ ORPs.  At the end of the simulation, the disc has not fragmented despite initially being in a critical state because with interstellar opacities, it heats up.}
  \label{fig:disc_Qmin0.75}
\end{figure}

\begin{figure}
  \includegraphics[width=1.0\columnwidth]{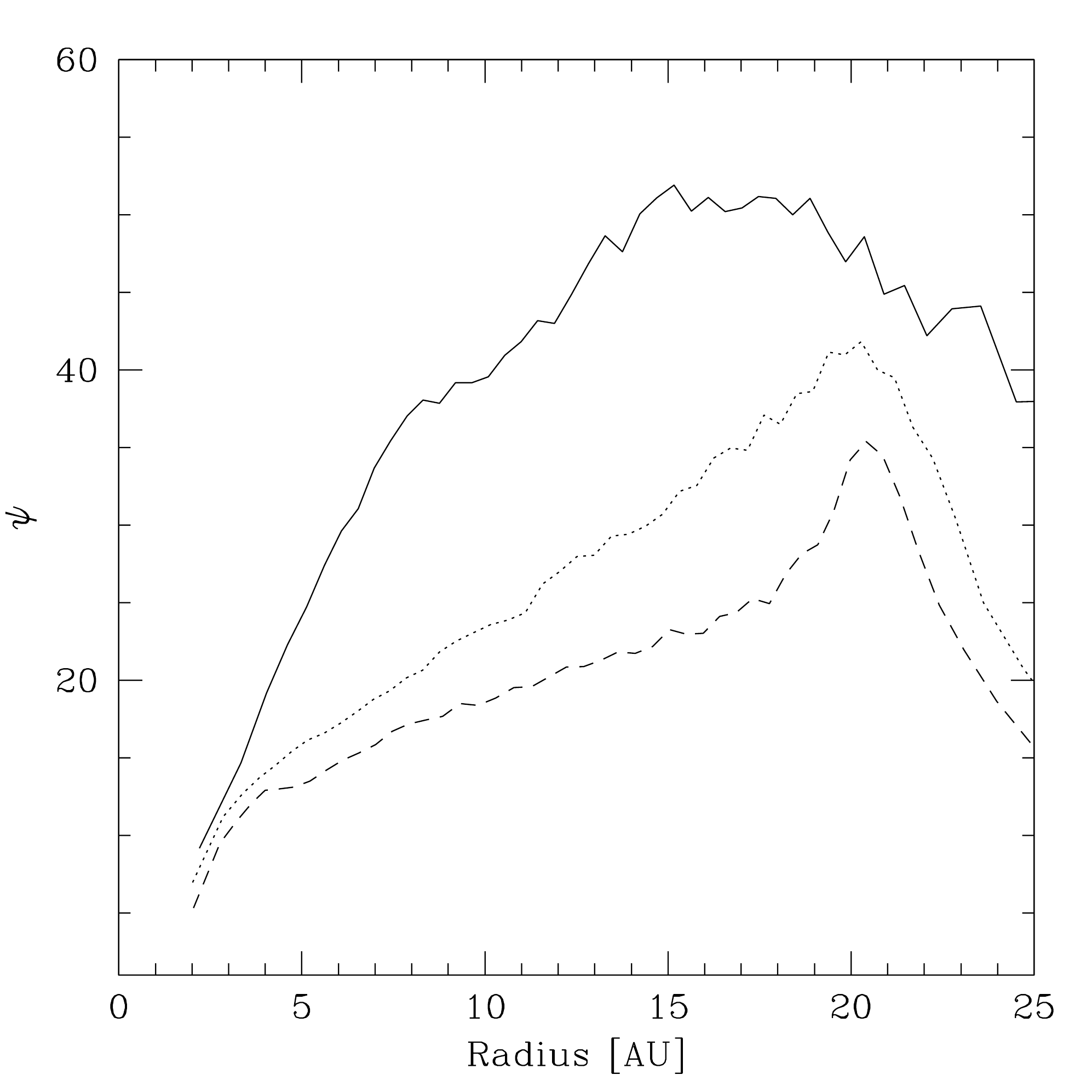}
  \caption{Cooling timescale, $\psi$, profiles for the Reference (solid line), Qmin1 (dotted line) and Qmin0.75 (dashed line) simulations at time $t = 6.4$ ORPs.  With interstellar opacities, these discs are simply not able to cool rapidly enough to obtain a $\psi$ value that is low enough for fragmentation.}
  \label{fig:Beta_Qmin}
\end{figure}

Figure~\ref{fig:Q_Qmin_start_end} shows the initial and final Toomre stability profiles for the Reference, Qmin1 and Qmin0.75 simulations.  The interesting aspect about the Qmin0.75 and Qmin1 cases are that though our initial and boundary conditions are either unstable or marginally stable, the discs still do not fragment because they are unable to cool rapidly and heat up so that they end up being at or above the marginal state (e.g. Figure~\ref{fig:disc_Qmin0.75}).  The cooling timescales, $\psi$, for the discs are as low as $\approx 20$ (Figure~\ref{fig:Beta_Qmin}) suggesting that for the discs to fragment, an efficient energy removing mechanism is needed such that the cooling timescale is lower than the Qmin0.75 (dashed line) curve in Figure~\ref{fig:Beta_Qmin}.

For completeness, we also simulate a disc with $Q_{\rm min} =~0.5$ and find that though this disc also heats up, it cannot heat fast enough to become Toomre stable before it fragments (due to its initial conditions).

\subsection{Low temperature, low opacity discs}

\begin{figure}
  \includegraphics[width=1.0\columnwidth]{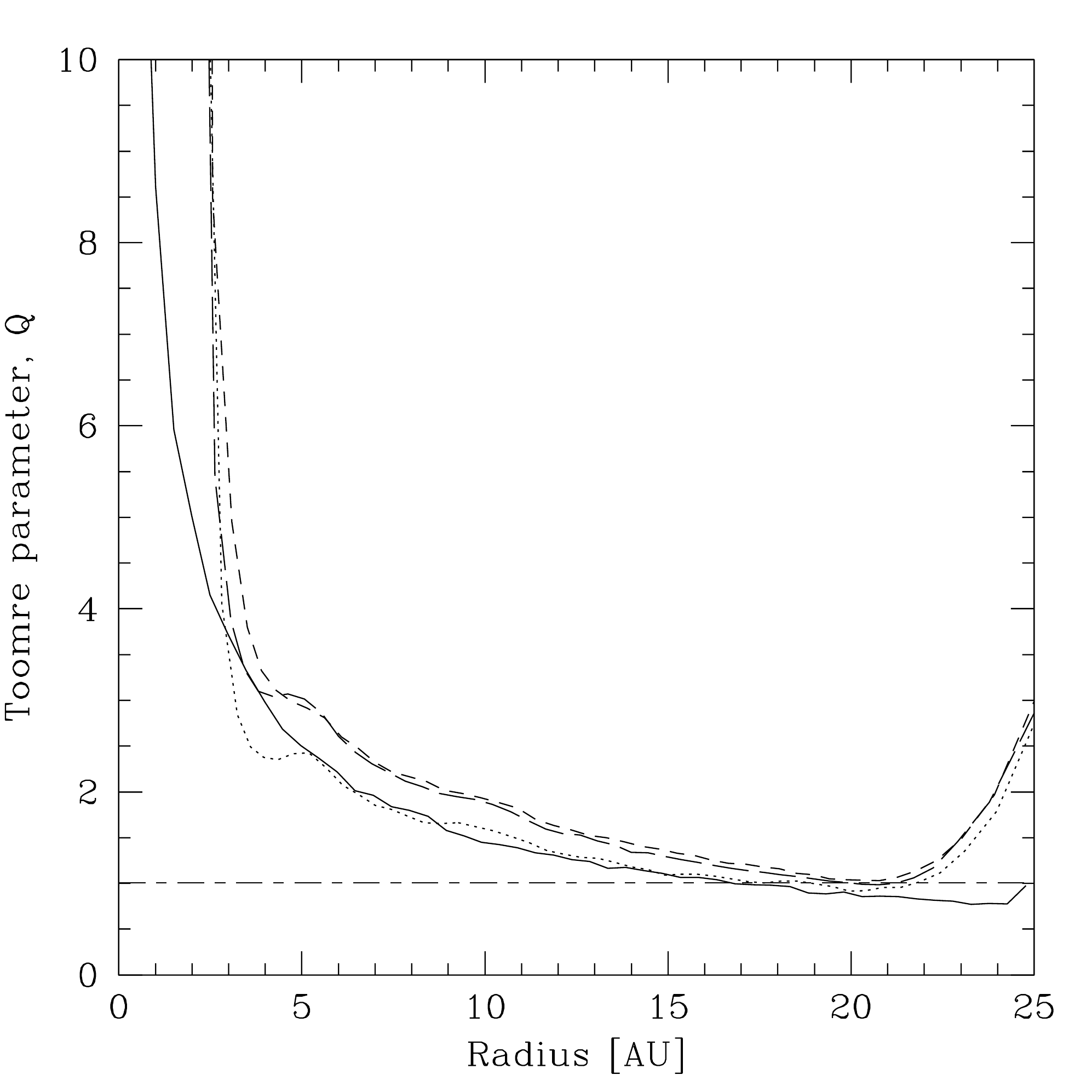}
  \caption{Toomre stability profiles at $t = 6.4$~ORPs for the Qmin0.75 (short dashed line) and Qmin0.75-Kappa0.1 (long dashed line) simulations and at $t = 8$~ORPs (just before it begins to fragment) for the Qmin0.75-Kappa0.01 (dotted line) simulation in comparison to the initial (solid line) Toomre stability profile for these discs.  The critical value of $Q_{\rm crit} = 1$ is also marked.  The lower opacity disc cools rapidly enough to attain a state of thermal equilibrium with its boundary.  It eventually fragments at $t = 9.7$~ORPs}
  \label{fig:Q_Qmin0.75_0.01IS}
\end{figure}

\begin{figure}
  \includegraphics[width=1.0\columnwidth]{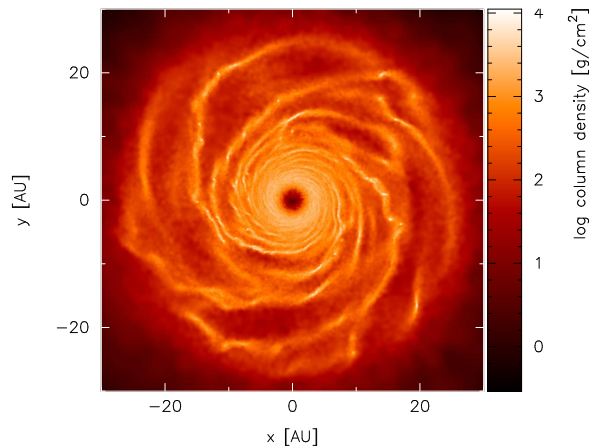}
  \caption{Surface density rendered image of the fragmented Qmin0.75-Kappa0.01 disc at a time of $t = 10.5$ ORPs.  The disc not only requires reduced irradiation, but also low opacities are essential for it to cool rapidly enough to fragment.}
  \label{fig:disc_small_frag}
\end{figure}

\begin{figure}
  \includegraphics[width=1.0\columnwidth]{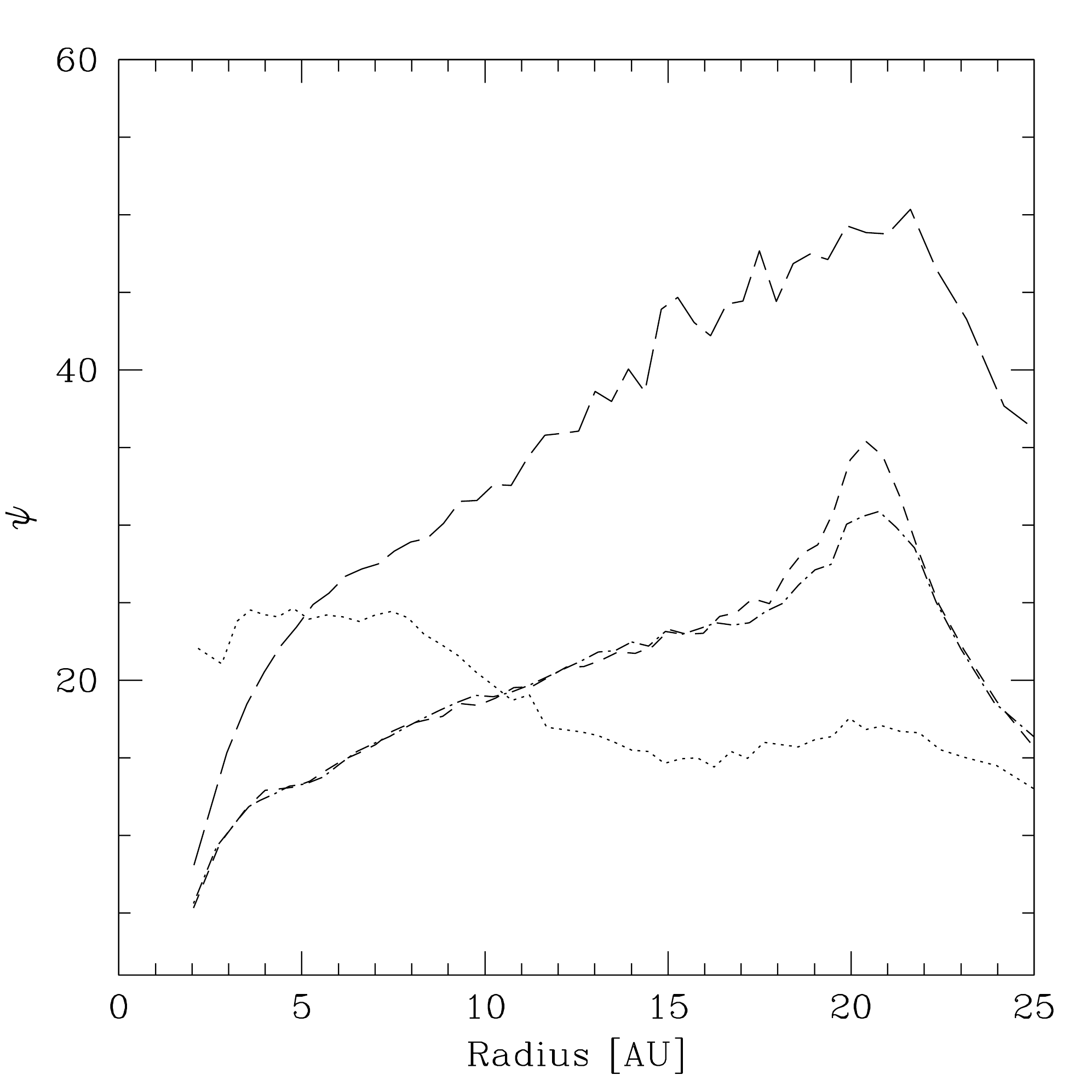}
  \caption{Cooling timescale, $\psi$, profiles for the Qmin0.75-Kappa0.01 simulation (dotted line) at $t = 8$ ORPs, just before it begins to fragment, in comparison to the Qmin0.75 (short dashed line), Kappa0.01 (long dashed line) and Qmin0.75-Kappa0.1 (dot-dashed line) simulations at $t = 6.4$ ORPs.  The fragmenting disc has a low $\psi$~value ($\approx 15$) in the outer part of the disc where it fragments.}
  \label{fig:Beta_small_frag_discs}
\end{figure}

Until now, we have identified under what conditions discs are \emph{more likely} to fragment (i.e. sub-interstellar opacities and cooler temperatures).  Combining these conditions, we simulate two further discs with outer Toomre stability parameters of 0.75 with $0.1\times$ and $0.01\times$ interstellar opacity values (simulations Qmin0.75-Kappa0.1 and Qmin0.75-Kappa0.01, respectively).  With such low opacity values, the Qmin0.75-Kappa0.01 disc would be equivalent to a low metallicity disc or a disc with grain sizes ranging between millimetre and centimetre sizes which is realistic given current observations \citep{Calvet_etal_cm_grains,Testi_etal_cm_grains,Rodman_etal_mm_grains,Lommen_etal_mm_grains}.

We can see from Figure~\ref{fig:Q_Qmin0.75_0.01IS} that the disc with $0.1 \times$ interstellar opacity values is slightly cooler than the case with interstellar opacities, but is still unable to cool rapidly enough to maintain a state of thermal equilibrium with its boundary.  However, the disc with $0.01 \times$ interstellar opacity values is able to since its Toomre stability profile at the end of the simulation is close to its initial value.  We find that the disc in the latter simulation does indeed fragment (Figure~\ref{fig:disc_small_frag}), though it takes $\sim9.7$ ORPs to do so.  This is because:

(i) the disc goes through a transient phase where it heats up since the initial disc is not quite in hydrostatic equilibrium.

(ii) although the radiation is able to leave quickly, the disc is enveloped in a \emph{thermal blanket} due to the effects of stellar irradiation, thus causing the disc to cool more slowly.

(iii) as the disc midplane cools, its cooling rate also decreases since the temperature gradient between the disc midplane and surface becomes smaller.

Figure~\ref{fig:Beta_small_frag_discs} shows the cooling timescale, $\psi$, of simulation Qmin0.75-Kappa0.01 (1.7~ORPs prior to fragmentation) in comparison to simulations Kappa0.01 (which has $Q_{\rm min}=2$), Qmin0.75 (which uses interstellar opacities) and Qmin0.75-Kappa0.1.  With a reduced opacity, the Qmin0.75-Kappa0.01 disc cools on a timescale fast enough such that it fragments i.e. it cools fast enough so that the heating due to gravitational instabilities, numerical viscosity and stellar irradiation do not heat the disc significantly above the boundary temperature.  This figure shows that the cooling timescale $\psi \approx 15$ in the outer parts of the disc, where fragmentation occurs.  We emphasise the need to express caution when considering the absolute value of $\psi$ in a non-steady state fragmenting disc, since it is dependent on the disc temperature which is constantly changing.  We discuss this in detail in Section~\ref{sec:disc}.  A key interesting aspect about this simulation is that the fragments form even though the disc is small ($\rm R < 25 AU$), a result contrary to what has been suggested in the past (see Section~\ref{sec:disc} for further discussion).

\subsection{300 AU discs}

\begin{figure}
  \includegraphics[width=1.0\columnwidth]{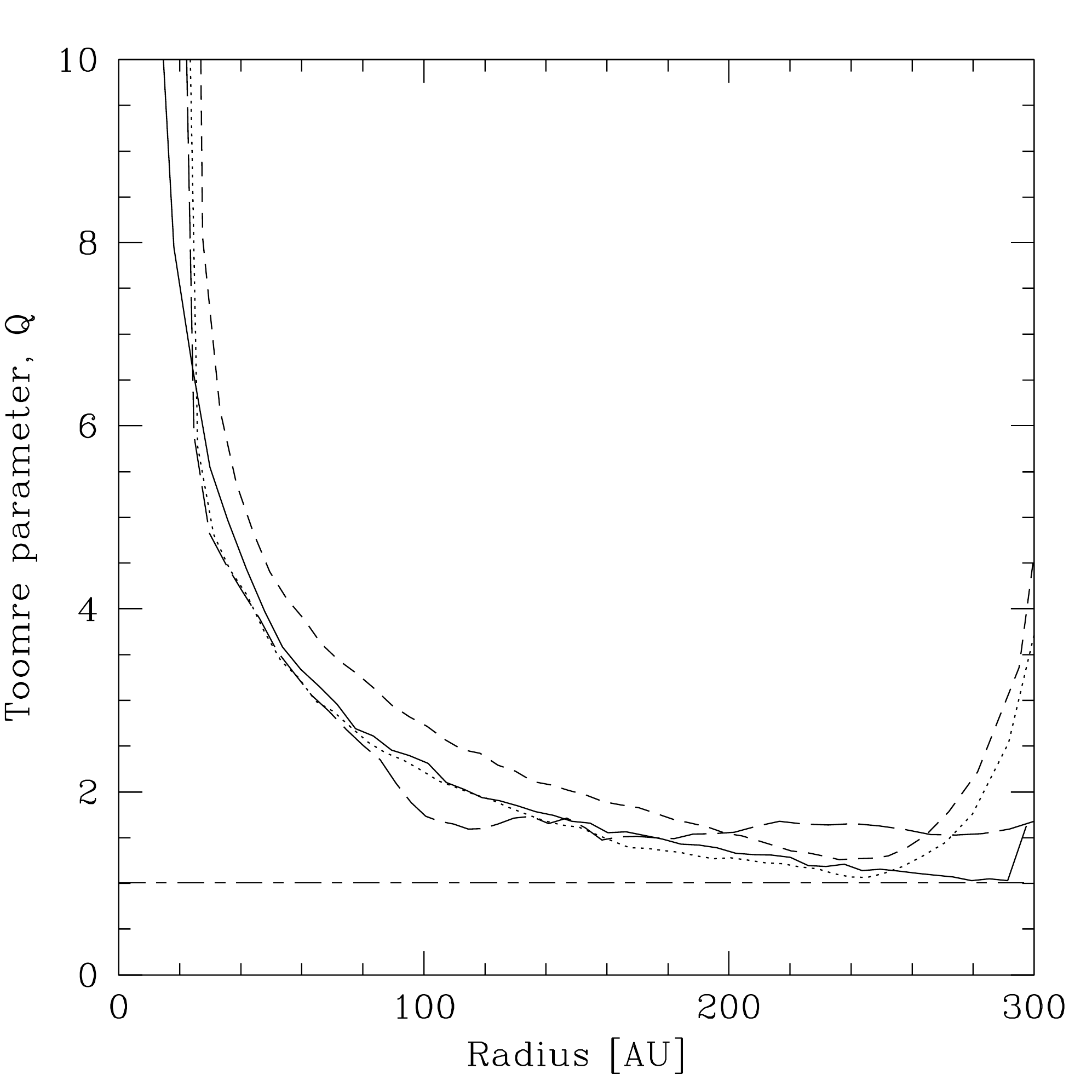}
  \caption{Toomre stability profiles for simulations L-Qmin1 (dotted line), L-Qmin1-Kappa10 (short dashed line) and L-Qmin1-Kappa0.1 (long dashed line) at a time of $t = 6.4$ ORPs as well as the boundary profile for these discs (solid line).  The critical value of $Q_{\rm crit} = 1$ is also marked.  Larger discs do not require as low opacities as smaller discs to attain a state of thermal equilibrium with their boundaries.}
  \label{fig:Q_Qmin1_kappa}
\end{figure}

\begin{figure}
  \includegraphics[width=1.0\columnwidth]{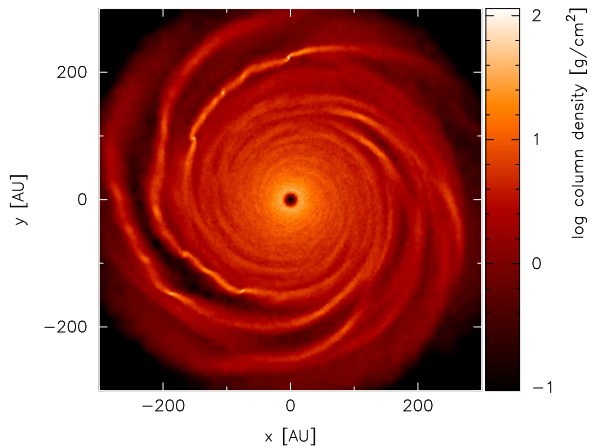}
  \caption{Surface density rendered image of the large, low opacity disc, L-Qmin1-Kappa0.1, at a time of $t = 2.9$ ORPs.  A low opacity is also required for a large disc to fragment, though the opacity does not have to be as low as the 25AU discs.}
  \label{fig:large_disc_fragment}
\end{figure}

We now turn our attention to larger discs that are usually considered more likely to promote fragmentation.

Simulations L-Qmin1-Kappa10, L-Qmin1 and L-Qmin1-Kappa0.1 show that though it is easier for larger discs to maintain a state of thermal equilibrium with their boundaries (Figure~\ref{fig:Q_Qmin1_kappa}), higher opacity discs still struggle to do so, thus implying that low opacities are still important for discs to fragment.

However, despite simulating a large disc with a marginal Toomre stability profile ($Q_{\rm min} = 1$; simulation L-Qmin1), we are still unable to see any fragmentation for discs with interstellar opacities unless we simulate discs whose fragmentation is determined by the initial conditions (e.g. Simulation L-Qmin0.75 which fragments in $t < 1$~ORP).  We find that though the opacities do not have to be as small as those in the 25AU discs ($0.01\times$ interstellar values), nevertheless, we still require discs with sub-interstellar opacities ($0.1\times$ interstellar values).  Figure~\ref{fig:large_disc_fragment} shows the disc of simulation L-Qmin1-Kappa0.1 which fragmented after more than an outer rotation period of evolving.  Though for fragmentation to occur, the opacity in this disc does not have to be decreased as much as for the smaller disc simulated in Qmin0.75-Kappa0.01, since it is colder, the grain sizes still correspond to sizes ranging between millimetre and centimetre which, as mentioned earlier, is entirely realistic.

\section{Comparison with previous work}
\label{sec:disc}

Our simulations show that contrary to past studies, it is possible for discs to fragment at small radii ($< 25$AU) if the disc temperature and the opacity are low enough.  The latter point may be the case if the disc is metal-poor or if grain growth has occurred to produce grains that are larger than interstellar sizes.  The larger grain size is certainly a reasonable assumption since there is evidence for grains of up to centimetre sizes in discs \citep{Calvet_etal_cm_grains,Testi_etal_cm_grains,Rodman_etal_mm_grains,Lommen_etal_mm_grains}, which could provide the low opacities necessary for our fragmenting discs.  With respect to disc metallicity, it is well known that core accretion is not as efficient in metal-poor environments \citep{Kornet_etal_CA_metallicity}.  However, we find that gravitational instability is enhanced in such conditions.  \cite{Boss_metallicity} carried out simulations of gravitationally unstable discs with varying opacities (from 0.1$\times$ to 10$\times$ the interstellar Rosseland mean values)  in order to explore fragmentation in different metallicity discs and found that the fragmentation results were insensitive to the dust grain opacity.  He also found that the disc midplanes could radiate energy to the disc surfaces very rapidly as the timescale for temperature equilibration by radiative diffusion between the disc midplane and the surface was smaller than the orbital timescale regardless of the opacity scaling.  However, for our discs the converse is true and we find that the opacity, and hence metallicity, does play a part in the likelihood of fragmentation.  Our results suggest that gravitational instability may be the dominant giant planet formation mechanism in low metallicity discs.

Our small fragmenting discs are also in contrast to previous work which has suggested via simulations \citep[e.g.][]{Stamatellos_no_frag_inside_40AU, Boley_CA_and_GI} and analytical work \citep[e.g.][]{Rafikov_unrealistic_conditions,Rafikov_SI,Clarke2009_analytical} that fragmentation at such small radii is not possible.  However, this is due to the lower opacities in our discs.  \cite{Rafikov_SI} analytically explored self-gravitating discs including the effects of stellar irradiation and suggested that fragmentation inside of $\approx 120 \rm{AU}$ is not possible.  However, firstly he uses interstellar Rosseland mean opacities (which is not an unreasonable assumption to initially make) and secondly he assumes that the fragmentation boundary occurs when the ``effective $\alpha$ parameter'' (of the form of \citealp{SS_viscosity}) due to the gravitational torques $\alpha_{\rm GI} \sim 1$.  Other authors \citep[e.g.][]{Clarke2009_analytical} assume the fragmentation boundary occurs when $\alpha_{\rm GI} \sim 0.06$ \citep{Gammie_betacool,Rice_beta_condition} which occurs at a radius $\rm R \sim 70$~AU.  This is still at a larger radii than our small fragmenting disc, though \cite{Clarke2009_analytical} also assumes interstellar Rosseland mean opacities.  Equations A1-A5 of \cite{Clarke2009_analytical} show that if the opacity is decreased by two orders of magnitude, as is the case for our small fragmenting disc, and using $\alpha_{GI} \sim 0.06$ as the value at which fragmentation can occur, the radius outside of which fragmentation occurs is at $R \approx 24$~AU.  This is much smaller than the canonical value of $\sim 70$~AU, although still not quite as small as $R \approx 15$~AU at which we find fragmentation.  Given that the work of \cite{Clarke2009_analytical} was analytical and while we have performed direct global three-dimensional radiation hydrodynamical simulations, this level of agreement is reasonable.

We can also compare our results to the analytical work of \cite{Rafikov_unrealistic_conditions}.  Based on a combination of the stability and cooling criteria of \cite{Toomre_stability1964} and \cite{Gammie_betacool}, respectively, \cite{Rafikov_unrealistic_conditions} analytically derived a constraint on the disc temperature that is required for planet formation via gravitational instability.  Using this constraint (equation 5 of \citealp{Rafikov_unrealistic_conditions}) and taking into account the reduced opacity in our discs, the Rafikov fragmentation conditions for the surface mass density and temperature in the disc become

\begin{equation}
\Sigma \ge \frac{\Omega^\frac{11}{10}}{Q_{crit} \pi G} \bigg(\frac{k}{\mu}\bigg)^{\frac{3}{5}}\bigg[\frac{1}{\zeta \sigma \kappa_o}\bigg]^{\frac{1}{10}},
\end{equation}
and

\begin{equation}
T \ge \bigg[\bigg(\frac{k}{\mu}\bigg)\frac{\Omega}{\zeta \sigma \kappa_o}\bigg]^{\frac{1}{5}},
\end{equation}
respectively, where $k$ is the Boltzmann constant, $\mu$ is the mean particle mass, $\sigma$ is the Stefan-Boltzmann constant, $\kappa_o$ is the constant of proportionality for the Rosseland mean opacity expression for this temperature and low opacity regime given by $\kappa = \kappa_o T^2$ and has a value $\kappa_o~=~2 \times 10^{-6} \rm g~cm^{-2}~K^{-2}$, and $\zeta = 2\beta(\gamma-1)$ and also absorbs any O(1) factors that have not been accurately considered in a proper calculation of the cooling time.  \cite{Rafikov_unrealistic_conditions} assumes that $\zeta \sim 1$.  We find that our fragmenting disc is in agreement with these conditions to with a factor of $\sim 2$.  As with the \cite{Clarke2009_analytical} comparison, this level of agreement is reasonable given that our simulations are global, three-dimensional and use a realistic radiative transfer method.  Consequently, our simulation results are consistent with previous analytical work, but they emphasise the importance of opacity in determining the radius outside of which fragmentation may occur.

Another difference between our simulations and those in the past that have used simplified cooling is that the heating in previous simulations has been dominated by internal heating processes i.e. heating due to gravitational instability and viscous processes, whereas our simulations involve additional external heating due to stellar irradiation.  This additional \emph{thermal blanket} causes the Toomre stability parameter to remain high, thus inhibiting fragmentation (consistent with the results of \citealp{Cai_envelope_irradiation} and \citealp{Stamatellos_no_frag_inside_40AU}).  However, in scenarios where the stellar irradiation is not so strong such that the Toomre stability parameter is small, fragmentation is possible, but only if the opacity is also decreased such that the disc can cool easily.

The cooling timescale, $\psi$, parameter cannot give any indication as to what the gravitational stress in a disc is since it incorporates the cooling in response to the heating due to stellar irradiation as well as the gravitational stresses.  However, it is still interesting to compare the values in our discs to the critical values of $\beta$ for fragmentation obtained from simulations using more simplified energetics.  Previous simulations using simplified cooling have suggested that for discs with $\gamma = 5/3$, fragmentation requires that the total cooling timescale in units of the orbital timescale, $\beta \lesssim 7$ \citep{Rice_beta_condition}.  Our non-fragmenting discs which have reached a steady state are consistent with this result since for these discs $\psi > 7$.  For our fragmenting disc, we find that the cooling timescale in units of the orbital timescale can be as much as $\psi \approx 15$.  However, we express caution when interpreting this result.  It is important to note that the \cite{Gammie_betacool} and \cite{Rice_beta_condition} conditions indicate the minimum cooling time (and hence a maximum gravitational stress) that a disc can support \emph{without} fragmenting.  In our fragmenting disc, the temperature continues to change as the disc cools, and hence the cooling rate is not constant since it is temperature dependent.

\section{Conclusions}
\label{sec:conc}
We have carried out radiation hydrodynamical simulations to investigate the evolution of massive self-gravitating discs ($M_{\rm disc}/M_{\star} = 0.1$).  We consider discs with opacities ranging from $0.01\times$ to $10\times$ the interstellar Rosseland mean values.  We also consider the effects of changing the initial and boundary temperatures of the discs as well as simulating different disc sizes (with outer disc radii, $R_{\rm out} = 25$ and 300 AU).

We find that the disc opacity is very important in determining whether a disc is likely to fragment.  In particular, we find that fragmentation is promoted in discs with opacity values lower than interstellar Rosseland mean values since this allows radiation to leave the disc quickly.  Low opacities may exist in low metallicity discs or discs with larger grain sizes.  This is a particularly important and timely result given the recent discoveries of wide orbit planets \citep{Fomalhaut,HR8799} and the future emphasis for surveys of planets on such wide orbits.  We show that it is possible for fragmentation to occur in gravitationally unstable discs even at radii where the innermost planet of the HR~8799 system is located ($R \gtrsim 24$~AU).  Furthermore, HR~8799 is known to be a metal-poor, $\lambda$ Bootis star with metallicity $[M/H] = -0.47$ \citep{HR8799_metallicity} so it is reasonable to assume that its disc was similarly metal-poor.  We have shown that such a scenario favours fragmentation and therefore, our results indicate that all three planets of the HR~8799 system may well have formed via gravitational instability.  Though a hybrid core accretion and gravitational instability scenario for planet formation may also be a possibility for this system, our calculations show that such a hybrid scenario may not be necessary.

We find that the presence of a \emph{thermal blanket} as a result of the stellar irradiation inhibits fragmentation since the discs are only able to cool to the boundary temperature.  However, we also show that under certain circumstances, fragmentation may occur.  Our results demonstrate that for fragmentation, weak irradiation is required such that the boundary temperature and hence Toomre stability parameter is low, in addition to low enough opacities (even in large, cool discs) since this allows more efficient cooling so that the disc's temperature does not increase significantly (due to internal dissipation) above the boundary temperature.

\section{Acknowledgments}
We thank the anonymous referee whose comments greatly improved the clarity of the paper.  The calculations reported here were performed using the University of Exeter's SGI Altix ICE 8200 supercomputer.  The disc images were produced using SPLASH \citep{SPLASH}.  MRB is grateful for the support of a EURYI Award which also funded FM. This work, conducted as part of the award `The formation of stars and planets: Radiation hydrodynamical and magnetohydrodynamical simulations' made under the European Heads of Research Councils and European Science Foundation EURYI (European Young Investigator) Awards scheme, was supported by funds from the participating organizations of EURYI and the EC Sixth Framework Programme.

\bibliographystyle{mn2e}
\bibliography{allpapers}

\end{document}